\documentclass[conference]{IEEEtran}
\IEEEoverridecommandlockouts
\usepackage{cite}
\usepackage{amsmath,amssymb,amsfonts}
\usepackage[ruled,linesnumbered,lined]{algorithm2e}
\usepackage{graphicx}
\usepackage{textcomp}
\usepackage{xcolor}
\usepackage{bm}
\def\BibTeX{{\rm B\kern-.05em{\sc i\kern-.025em b}\kern-.08em
    T\kern-.1667em\lower.7ex\hbox{E}\kern-.125emX}}

\begin{document}
	\title{Reconfigurable Intelligent Surfaces Empowered Cooperative Rate Splitting with User Relaying}
	\author{
		\IEEEauthorblockN{Kangchun Zhao$^{\star}$, Yijie Mao$^{\star}$, Zhaohui Yang$^{\dagger}$, Lixiang Lian$^{\star}$, and  Bruno Clerckx$^{\ddagger \star}$ }
		\IEEEauthorblockA{
		$^{\star}$School of Information Science and Technology, ShanghaiTech University, Shanghai 201210, China \\
		$^{\dagger}$College of Information Science and Electronic Engineering, Zhejiang University, Hangzhou, Zhejiang 310027, China \\
		$^{\ddagger \star}$Department of Electrical and Electronic Engineering,	Imperial College London, United Kingdom\\
			Email:13238003365@163.com, maoyj@shanghaitech.edu.cn, yang\_zhaohui@zju.edu.cn, lianlx@@shanghaitech.edu.cn,\\ b.clerckx@imperial.ac.uk}
		\thanks{This work has been partially supported by  partially supported by Shanghai Sailing Program under Grant 22YF1428400.}
		\\[-1.5 ex]
		{\sublargesize\textit{(Invited Paper)}}	
		\\[-3.5 ex]		
	}
\maketitle
\thispagestyle{empty}
\pagestyle{empty}
\begin{abstract}
Cooperative rate splitting (CRS), built upon rate splitting multiple access (RSMA) and opportunistic user relaying, has been recognized as a promising transmission strategy to enhance the user fairness and spectral efficiency in multi-antenna broadcast channels. 
To further boost its performance, the interplay of CRS and reconfigurable intelligent surface (RIS) is investigated in this work. 
Specifically, a novel RIS-aided CRS transmission framework is  proposed and the corresponding resource allocation problem to maximize the minimum rate among users is investigated.  
An alternative optimization algorithm is then proposed to optimize the transmit beamforming, common rate allocation, and RIS phases, iteratively. 
Numerical results show that the proposed RIS-aided CRS transmission framework significantly improves the spectral efficiency compared with its non-cooperative counterpart and  other schemes without RIS. 
\end{abstract}


\section{Introduction}
Among numerous potential techniques for the sixth generation (6G) communication networks,  rate-splitting multiple access (RSMA), as a novel non-orthogonal transmission framework and interference management strategy in the physical (PHY) layer,  has shown its great potential for improving the spectral efficiency, energy efficiency, user fairness, robustness to the channel state information uncertainty, etc \cite{mao2022rate}. 
The principle of (1-layer) RSMA is to split user messages into common and  private parts, encode the common parts into a common stream using a codebook shared by all users while encoding the private parts independently for the corresponding users only. By superposing the common stream on top of the private streams at the transmitter and  decoding  the common stream and the intended private stream sequentially at the receivers, RSMA achieves a more versatile interference management of partially decoding the interference and partially treating the interference as noise \cite{RSintro16bruno}.
Such 1-layer RSMA only requires one layer of successive interference cancellation (SIC) at each receiver, and is a more generalized transmission scheme than orthogonal multiple access (OMA) and  linearly-precoded space division multiple access (SDMA) \cite{mao2017rate,bruno2019wcl}.
One major characteristic of 1-layer RSMA is that the common stream is required to be decoded by all users and therefore the achievable common rate\footnote{The rate of the common stream is simply denoted as ``common rate".} is limited by the worst-case rate of decoding the common stream at all users. 
To enhance the common rate, a two-user cooperative rate splitting (CRS) is proposed in  \cite{jian2019crs}, and further extended to the $K$-user case in \cite{Mao2020}.
By enabling one user to opportunistically forward its decoded common message to the user with the worst-case common rate, CRS enlarges the rate region \cite{jian2019crs}, enhances user fairness \cite{Mao2020}, improves the coverage \cite{mao2022rate}, and maximizes the secrecy rate \cite{Liping2020secrecyCRS} compared with cooperative non-orthogonal multiple access (NOMA) and non-cooperative 1-layer RSMA, SDMA, NOMA. 

\par
Another enabling technique for 6G that has gained significant attention is reconfigurable intelligent surface (RIS) \cite{huang2018achievable}.
As a meta-surface containing a large number of discrete and passive elements, RIS is capable of  controlling wireless channels  by dynamically reconfiguring the reflection coefficients. Therefore, the spectral and energy efficiencies can be further improved. 
Inspired by the appealing performance benefits of RIS and RSMA, existing works have investigated the interplay between RIS and RSMA in terms of the outage probability \cite{bansal2021rate,bansal2021analysis}, max-min fairness \cite{fu2021resource}, and spectral efficiency \cite{fang2022fully, hongyu2022wcl}. However, all the above works only consider RIS-aided non-cooperative RSMA. So far, the interplay of CRS and RIS has not been investigated yet. 
\par
In this paper, we propose a novel RIS-aided CRS transmission framework, which enables an RIS to assist the direct transmission from the base station (BS) to the users and the opportunistic transmission of the common stream between users. 
We formulate  a max-min fairness optimization problem to jointly optimize RIS phases, transmit beamforming, common rate allocation, and the time slot allocation between the direct  and opportunistic transmissions under the transmit power constraint and the unit-modulus constraints of the RIS phases.
To solve the problem,  an alternative optimization (AO) algorithm is proposed to iteratively optimize the transmit beamforming and the RIS phases. 
Numerical  results show that the proposed RIS-aided CRS transmission framework improves the worst-case achievable rate among users compared with its
non-cooperative counterpart and other schemes without RIS. Therefore, we conclude that the proposed RIS-aided CRS scheme enhances user fairness and is more powerful than the existing transmission schemes. 

\section{System Model and Problem Formulation}
\label{sec: system model}
\subsection{System Model}
Consider a multi-user multiple-input single-output (MISO) transmission network, where a BS equipped with $N_t$ transmit antennas simultaneously serves two single-antenna users indexed by $U_1$ and $U_2$.
Without loss of generality, $U_1$ is assumed to have a better channel to the BS than $U_2$, and it opportunistically acts as a half-duplex (HD) relay to forward the signal to $U_2$. 
There is one RIS with $N$ passive reflecting elements to assist the transmission from BS to the two users and from $U_1$ to $U_2$. The RIS elements are indexed by $\mathcal{N}=\{1,\ldots,N\}$.
 The channels between BS and RIS, BS and $U_k$, RIS and $U_k$, $U_1$ and $U_2$ are denoted by $\mathbf{G}\,\in\,\mathbb{C}^{N\times N_t},\,\mathbf{g}_k\,\in\,\mathbb{C}^{N_t\times 1}$, $\mathbf{h}_{k}\,\in\,\mathbb{C}^{N\times1}$ and $h_{12}$, respectively. The system model is delineated in Fig. 1.
\par 
Two phases are involved in the proposed model, namely, the direct transmission phase and the cooperative transmission phase. 
In the direct transmission phase, also known as (a.k.a.) the first time slot, the BS transmits signals to two users based on the principle of 1-layer RSMA. In the meanwhile, the transmit signal is reflected by the RIS to the two users.
In the cooperative transmission phase, a.k.a. the second time slot, the BS is silent and 
$U_1$  transmits the common stream of 1-layer RS to $U_2$ and the signal is reflected by the RIS to $U_2$. 
We assume $\beta\enspace (0< \beta \leq 1)$ as the fraction of time allocated to the first phase.
The rest $(1-\beta)$ is allocated to the second phase.

\subsubsection{Direct Transmission Phase}
Following the principle of 1-layer RSMA  \cite{RS2016hamdi}, the message $W_{k}$ intended to user-$k \in\{1,2\}$ is split into a common part $W_{c,k}$ and a private part $W_{p,k}$.
$W_{c,1}$ and $W_{c,2}$ are combined and encoded into common stream $s_0$ using a common codebook shared by the two users.
The private parts $W_{p,1}$ and $W_{p,2}$ are independently encoded into private streams $s_1$ and $s_2$, respectively. 
Assume each stream $s_k$ has zero mean and unit variance, i.e., $\mathbb{E}\{s_k^{H}s_k\}=1$.
Let $\mathbf{s}=[s_0,s_1,s_2]^T$ and $\mathbf{P}=[\mathbf{p}_0,\mathbf{p}_1,\mathbf{p}_2 ]\in \mathbb{C}^{N_t\times 3}$ respectively denote the stream vector and the beamforming matrix, where $\mathbf{p}_k\in\mathbb{C}^{N_t\times 1}$.
The resulting transmit signal at the BS in the first time slot is expressed as
\begin{equation}
 	\mathbf{x}^{(1)}=\mathbf{P}\mathbf{s}=\sum_{k=0}^{2}\mathbf{p}_ks_k.
 \end{equation}
The transmit power constraint is given by  $\textrm{tr}(\mathbf{P}\mathbf{P}^H)\leq P_t$, where $P_t$ is the maximum transmit power of the BS. 
The received signal at user-$k$ in the first time slot is given as
\begin{equation}
\label{2}
  y_{k}^{(1)}=(\mathbf{g}_k^H+\mathbf{h}_{k}^H\mathbf{\Theta}_{(1)}\mathbf{G})\mathbf{x}+n_k,\forall  k\in \{1,2\},
  \end{equation}
where  $\mathbf{\Theta}_{(1)}=\textrm{diag}(e^{j\theta_{1}^{(1)}},\ldots,e^{j\theta_N^{(1)}})\in\mathbb{C}^{N\times N}$  is the phase matrix in the first time slot.
$n_k$ is the additive white Gaussian noise (AWGN) which follows $n_k\sim  \mathcal{CN}(0, \sigma_k^{2})$.
\begin{figure}[tb]
\centerline{\includegraphics[width=0.35\textwidth]{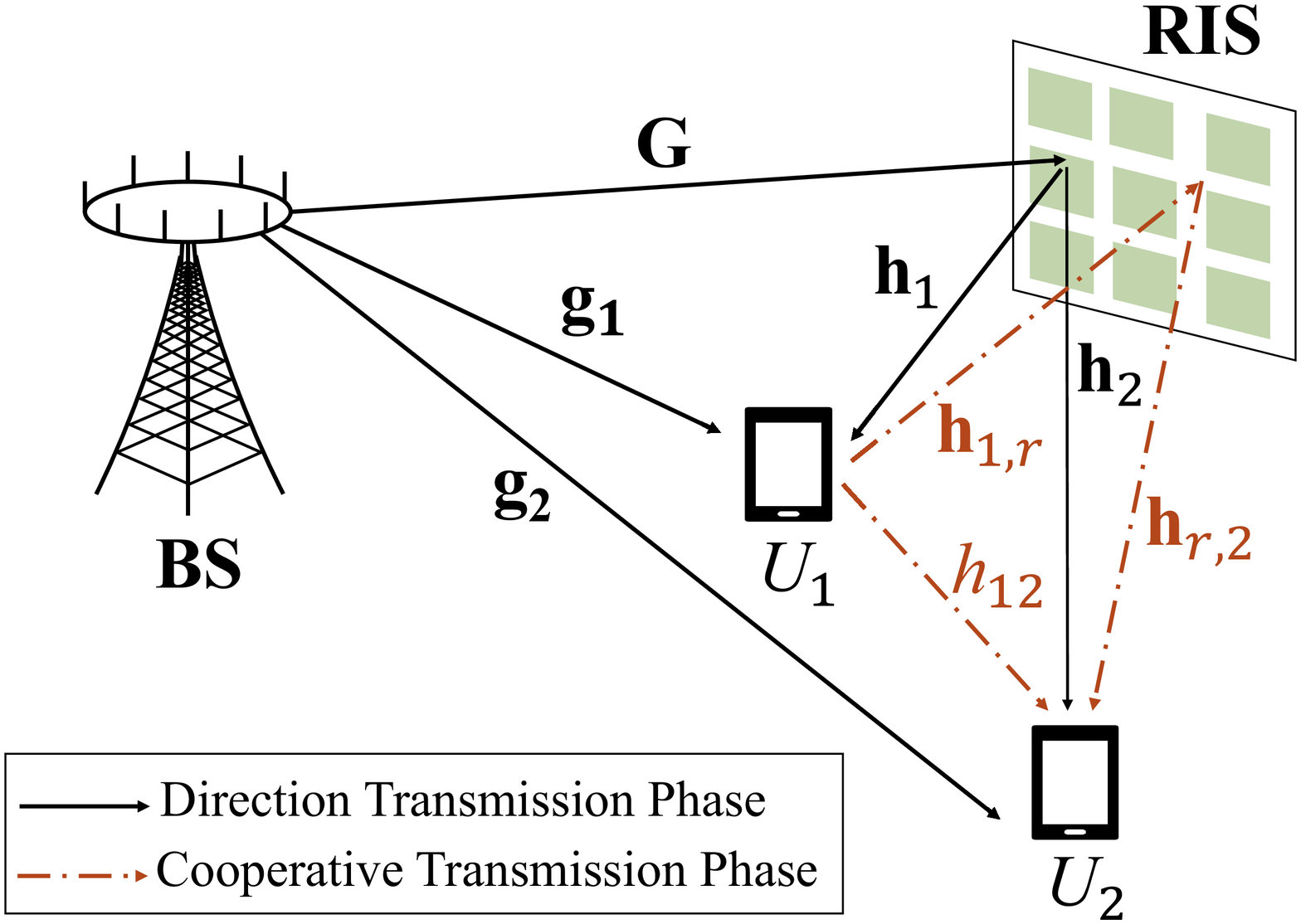}}
\vspace{-1mm}
\caption{The proposed RIS-aided cooperative rate splitting transmission architecture.}
\label{fig1}
\vspace{-3mm}
\end{figure}
Once user-$k$ receives $y_k^{(1)}$ in the first time slot, it first decodes the common stream $s_0$ by fully treating all private streams as interference.
The achievable rate of decoding the common stream at user-$k$ in the first time slot is 
\begin{equation}
c_{k}^{(1)}=\beta\log_2\left(1+\frac{\left|(\mathbf{g}_k^H+\mathbf{h}_{k}^H\mathbf{\Theta}_{(1)}\mathbf{G})\mathbf{p}_0\right|^2}{\sum_{i=1}^{2}\left|(\mathbf{g}_k^H+\mathbf{h}_{k}^H\mathbf{\Theta}_{(1)}\mathbf{G})\mathbf{p}_i\right|^2+\sigma_{k}^2}\right).
 \end{equation} 
After removing the decoded common stream, the rate of decoding the private stream is
\begin{equation}
 r_{k}^{(1)}=\beta\log_2\left(1+\frac{\left|(\mathbf{g}_k^H+\mathbf{h}_{k}^H\mathbf{\Theta}_{(1)}\mathbf{G})\mathbf{p}_k\right|^2}{\left|(\mathbf{g}_k^H+\mathbf{h}_{k}^H\mathbf{\Theta}_{(1)}\mathbf{G})\mathbf{p}_i\right|^2+\sigma_{k}^2}\right),
 \end{equation}
 where $k,i \in 1,2$ and $i\neq k$. 
 \subsubsection{Cooperative Transmission Phase}
 In the cooperative transmission phase,  the BS is silent and $U_1$ forwards the decoded common stream $s_0$ to $U_2$ by employing the non-regenerative decode-and-forward (NDF) protocol \cite{jian2019crs} with the transmit power $P_r$, a phase matrix $\bm{\Theta}_{(2)}$ and  a codebook generated independently at the BS.
 The received signal at $U_2$ with the aid of RIS  is given as
 \begin{equation}
y_{2}^{(2)} =(h_{12}+\mathbf{h}_{r,2}^H\mathbf{\Theta}_{(2)}\mathbf{h}_{1,r})\sqrt{P_{r}}s_0+n_2,
\end{equation}
where $\mathbf{h}_{1,r}$ and $\mathbf{h}_{r,2}$ are channels between $U_1$ and RIS, $U_2$ and RIS, respectively. The rate of decoding the common stream at $U_2$ in the second time slot is
 \begin{equation}
 c_{2}^{(2)}=(1-\beta)\log_2\left(1+\frac{P_{r}\left|h_{12}+\mathbf{h}_{r,2}^H\mathbf{\Theta}_{(2)}\mathbf{h}_{1,r}\right|^2}{\sigma_2^{2}}\right).
 \end{equation} 
$U_2$  combines the decoded common stream from two time slots.
To make sure the common stream $s_0$ can be successfully decoded by both users, the achievable rate of decoding the common stream $s_0$ at $U_1$ and $U_2$  is obtained as follows 
\begin{equation}
 \label{R_c}
 R_{c}=\min\left(c_{1}^{(1)},c_{2}^{(1)}+c_{2}^{(2)}\right).
  \end{equation}
$R_{c}$ is shared by both users for the transmission of $W_{c,1}$ and $W_{c,2}$. Accordingly, it satisfies
\begin{equation}
    a_1+a_2\leq R_c,
\end{equation}
where $a_1$ and $a_2$ are parts of $R_c$ allocated to transmit the common parts of the two users. 
Let $\mathbf{a}=[a_1,a_2]$.
The total achievable rate of user $k$ is $R_{k,tot}=r_{k}^{(1)} +a_k,\forall k \in \{1,2\}$.
\subsection{Problem Formulation}
The goal of this work is to maximize the user fairness.
In particular, we jointly design the transmit beamforming $\mathbf{P}$, rate allocation $\mathbf{a}$, RIS phases $\bm{\Theta}=[\bm{\Theta}_{(1)},\bm{\Theta}_{(2)}]$ and time slot allocation $\beta$ with the aim of maximizing the minimum rate (max-min rate) of the users while satisfying the transmit power constraint. 
The max-min  rate  problem for the  proposed RIS-aided CRS is formulated as
\begin{subequations}\label{eq: max-min rs}\begin{align}
(\mathcal{P}_1)\,\,\,\,&\max_{\mathbf{{P}}, \mathbf{a}, \bm{\Theta},\beta }\,\,\min_{k\in \{1,2 \}} \,\,R_{k,tot} \label{a}\\
\mbox{s.t.}\quad
	& a_1+a_2\leq R_c,\label{a1}\\
	& a_{k}\geq 0,\forall k\in \{1,2\},\label{a2}\\
	& \bm{\Theta}_{(m)}=\text{diag}(e^{j\theta_1^{(m)}},\ldots,e^{j\theta^{(m)}_N}),m\in\{1,2\},\label{a3}\\
	&\theta_n^{(m)}\in [0,2\pi] ,\forall n\in \mathcal{N},m\in\{1,2\},\label{a4}\\
	&\text{tr}(\mathbf{P}\mathbf{P}^{H})\leq P_{t}, \label{a5}\end{align}\end{subequations}
 where constraint (\ref{a1}) ensures that each user successfully decodes the common stream. 
 Constraint (\ref{a4}) is the range of the phase for each RIS element,  and (\ref{a5}) presents the transmit power constraint at the BS. 
 Problem $\mathcal{P}_1$ is highly intractable and non-convex. To solve the problem, we develop an AO-based optimization algorithm in the next section to alternatively optimize the RIS phases $\mathbf{\Theta}$ and the remaining variables $\mathbf{P,a}, \beta$.

\section{Proposed Optimization Framework}
In this section, we specify the proposed AO framework to solve problem $\mathcal{P}_1$. The problem is first decomposed into two subproblems, one for joint transmit beamforming, common rate and time slot optimization, and the other one for the RIS phase matrix optimization. Both subproblems
are respectively solved by  success convex approximation (SCA)-based algorithms in an iterative manner until convergence.
\vspace{-2mm}
\subsection{Joint Beamforming, Common Rate, and Time Slot Optimization}
With given $\bm{\Theta}$, the channels from BS to IRS and IRS to users are fixed. 
For notational simplicity, we denote the effective channel in the first time slot as $\mathbf{\widetilde{g}}_k=\mathbf{g}_k+\mathbf{G}^H\mathbf{\Theta}_{(1)}^{H}\mathbf{h}_k$.
By introducing an auxiliary variable $t$ to denote the minimum rate of the two users, problem $\mathcal{P}_1$ is equivalently transformed to 
\begin{subequations}
\label{precoder simple}
\begin{align}(\mathcal{P}_2)\,\,\,\,&\max_{\mathbf{{P}}, \mathbf{a},  \beta,t}\,\,\,\,\,\,\,t \label{eqv step2 const:o1}\\
	\mbox{s.t.}\quad
	& r_{k}^{(1)} +a_k\geq t, \forall k\in\{1,2\}, \label{ps:1}\\
	&\,\, \textrm{(\ref{a1}), (\ref{a2}), (\ref{a5})}. \nonumber
	\end{align}
\end{subequations}
$\mathcal{P}_2$ is still non-convex due to the rate expressions $r_k^{[1]}$ and $R_c$ in (\ref{ps:1}) and (\ref{a1}).
To solve $\mathcal{P}_2$, we introduce slack variable vectors $\bm{\alpha}=[\alpha_1,\alpha_2]$, $\bm{\alpha}_c=[\alpha_{c,1},\alpha_{c,2}]$, $\bm{\rho}=[\rho_1,\rho_2]$ and $\bm{\rho}_c=[\rho_{c,1},\rho_{c,2}]$, where 
$\bm{\alpha}$ and $\bm{\alpha}_c$ denote the private and common stream rate vectors, respectively, $\bm{\rho}$ and $\bm{\rho}_c$ denote the signal-to-interference-plus-noise ratio (SINR) vectors for private  and common streams, respectively. 
With the slack variables introduced above,  $\mathcal{P}_2$ is equivalently transformed into 
\vspace{-2mm}
\begin{subequations}
	\label{precoder hard}
	\begin{align}(\mathcal{P}_{2.1})\,\,\,\,&\max_{\substack{\mathbf{{P}}, \mathbf{a}, \beta,t, \\ \bm{\alpha},\bm{\alpha}_c,\bm{\rho},\bm{\rho}_c}}\,\,\,\,\, \,\,t \label{ph:1}\\
	\mbox{s.t.}\quad
	&\,\, \beta\alpha_k+a_k\geq t,\forall k\in\{1,2\},  \label{ph:2}\\
	&\,\, \beta\alpha_{c,1}\geq a_1+a_2, \label{ph:3}\\
	&\,\,  \beta\alpha_{c,2}+c_{2}^{(2)}\geq a_1+a_2,\label{ph:4}\\
	&\,\, 1+\rho_k-2^{\alpha_k}\geq 0,\forall k\in\{1,2\},  \label{ph:5}\\
	&\,\, 1+\rho_{c,k}-2^{\alpha_{c,k}}\geq 0,\forall k\in\{1,2\}, \label{ph:6}\\
	&\,\, \frac{\left|\mathbf{\widetilde{g}}_k^H\mathbf{p}_k\right|^2}{\left|\mathbf{\widetilde{g}}_k^H\mathbf{p}_i\right|^2+\sigma_k^2}\geq \rho_k,\forall k\in\{1,2\},i\neq k, \label{ph:7}\\
	&\,\, \frac{\left|\mathbf{\widetilde{g}}_k^H\mathbf{p}_0\right|^2}{\sum_{i=1}^{2}\left|\mathbf{\widetilde{g}}_k^H\mathbf{p}_i\right|^2+\sigma_k^2}\geq \rho_{c,k},\forall k\in\mathcal{K},\label{ph:8}\\
	&\,\, \textrm{(\ref{a2}), (\ref{a5})}. \nonumber
\end{align}
\end{subequations}
To handle the nonconvexity of constraints (\ref{ph:2})--(\ref{ph:4}), (\ref{ph:7}) and (\ref{ph:8}), we next apply the SCA method. Constraint (\ref{ph:2}) contains a bilinear function $\beta\alpha_k$, which can be rewritten as $\beta\alpha_k=\frac{1}{4}(\beta+\alpha_k)^{2}-\frac{1}{4}(\beta-\alpha_k)^{2}$. 
Therefore, $\beta\alpha_k$ can be approximated by the first-order Taylor approximation at the point $(\beta^{[l]} ,\alpha_k^{[l]})$, which is given as
\begin{equation}
\label{b}
\begin{aligned}
\beta\alpha_k \geq \frac{1}{2}(\beta^{[l]}+\alpha_k^{[l]})(\beta&+\alpha_k)-\frac{1}{4}(\beta^{[l]}+\alpha_k^{[l]})^2\\
&-\frac{1}{4}(\beta-\alpha_k)^2\triangleq  \Phi^{[l]}(\beta,\alpha_{k}). 
 \end{aligned}
\end{equation}
With (\ref{b}),  (\ref{ph:2})--(\ref{ph:4}) are approximated at iteration $l$ around the point $(\beta^{[l]},\bm{\alpha}^{[l]},\bm{\alpha}^{[l]}_c)$ as
\begin{equation}
\label{bilinear}
\begin{aligned}
&\,\, \Phi^{[l]}(\beta,\alpha_{k})+a_k\geq t,\forall k\in\{1,2\}, \\
&\,\, \Phi^{[l]}(\beta,\alpha_{c,1})\geq a_{1}+a_2, \\
&\,\,  \Phi^{[l]}(\beta,\alpha_{c,2})+c_{2}^{(2)}\geq a_1+a_2. 
\end{aligned}
\end{equation}
By further  transforming (\ref{ph:7}) and (\ref{ph:8}) into the following difference-of-convex (DC) forms
and using the first-order Taylor approximations to reconstruct the concave parts of the DC constraints,  constraints (\ref{ph:7}) and (\ref{ph:8}) at iteration $l$ are approximated respectively at the point $(\mathbf{P}^{[l]},\bm{\rho}^{[l]},\bm{\rho}_c^{[l]})$ by
\begin{equation}
\label{eq: DC approxi}
\begin{aligned}
|\mathbf{\widetilde{g}}_{k}^{H}\mathbf{{p}}_{i}|^{2}+\sigma^2_k&-\frac{2\Re\{(\mathbf{{p}}_{k}^{[l]})^H{\mathbf{\widetilde{g}}}_{k}{\mathbf{\widetilde{g}}}_{k}^{H}\mathbf{{p}}_{k}\}}{ \rho_k^{[l]}}+\frac{|\mathbf{\widetilde{g}}_{k}^{H}\mathbf{{p}}_{k}^{[l]}|^{2}\rho_k}{(\rho_k^{[l]})^2}\\&\quad\quad\quad\quad\quad\quad\quad\leq 0,\forall k\in\{1,2\},i\neq k,\\
\sum_{i=1}^{2}|\mathbf{\widetilde{g}}_{k}^{H}\mathbf{{p}}_{i}|^{2}+\sigma^2_k&-\frac{2\Re\{(\mathbf{{p}}_{0}^{[l]})^H{\mathbf{\widetilde{g}}}_{k}{\mathbf{\widetilde{g}}}_{k}^{H}\mathbf{{p}}_{0}\}}{ \rho_{c,k}^{[l]}}+\frac{|\mathbf{\widetilde{g}}_{k}^{H}\mathbf{{p}}_{0}^{[l]}|^{2}\rho_{c,k}}{(\rho_{c,k}^{[l]})^2}\\&\quad\quad\quad\quad\quad\quad\quad\leq 0,\forall k\in\{1,2\}.
\end{aligned}
\end{equation}
\begin{algorithm}	
 	\textbf{Initialize}: $n\leftarrow0,t^{[l]}\leftarrow0$, $\mathbf{{P}}^{[l]},\beta^{[l]}, \bm{\alpha}^{[l]},\bm{\alpha}_c^{[l]},\bm{\rho}^{[l]},\bm{\rho}_c^{[l]}$\;
 	\Repeat{$|t^{[l]}-t^{[l-1]}|<\epsilon$}{
 		$l\leftarrow l+1$\;
 Solve  problem (\ref{p-last}) using  $\mathbf{{P}}^{[l-1]},\beta^{[l-1]}, \bm{\alpha}^{[l-1]},$ $\bm{\alpha}_c^{[l-1]},\bm{\rho}^{[l-1]},\bm{\rho}_c^{[l-1]}$ and denote the optimal value of the objective function by $t^{\star}$ and the optimal solutions by $\mathbf{{P}}^{\star},\beta^{\star}, \bm{\alpha}^{\star},\bm{\alpha}_c^{\star},\bm{\rho}^{\star},\bm{\rho}_c^{\star}$\;
 Update $t^{[l]}\leftarrow t^{\star}$, $\mathbf{P}^{[l]}\leftarrow \mathbf{P}^{\star}$,  $\beta^{[l]}\leftarrow\beta^{\star}$, $\bm{\alpha}^{[l]}\leftarrow\bm{\alpha}^{\star}$, $\bm{\alpha}_c^{[l]}\leftarrow\bm{\alpha}_c^{\star}$, $\bm{\rho}^{[l]}\leftarrow\bm{\rho}^{\star}$, $\bm{\rho}_c^{[l]}\leftarrow\bm{\rho}_c^{\star}$\;			 	}	
\caption{Joint beamforming, common rate, and time slot optimization algorithm for problem $\mathcal{P}_2$ }
\end{algorithm}
With the above approximations,  problem $\mathcal{P}_{2}$ at iteration $l$ can be  approximated by the following convex problem
\begin{equation}
\label{p-last}
    \begin{aligned}
    \max_{\substack{\mathbf{{P}}, \mathbf{a}, \beta,t, \\ \bm{\alpha},\bm{\alpha}_c,\bm{\rho},\bm{\rho}_c}}\,\,\,\,&\, \,\,t\\
    	\mbox{s.t.}\quad
&\,\,  \textrm{(\ref{a2}), (\ref{a5}), (\ref{ph:5}), (\ref{ph:6}), (\ref{bilinear}), (\ref{eq: DC approxi})}.  	
 \end{aligned}
\end{equation}
Problem (\ref{p-last}) is a convex problem, which can be solved using the SCA method. The detailed process of the SCA method to solve $\mathcal{P}_2$ is illustrated in Algorithm 1.
\subsection{RIS Phase Optimization}
Given transmit beamforming vector $\mathbf{P}$, rate allocation $\mathbf{a}$, and time slot allocation $\beta$, we also introduce an auxiliary variable $t$ and reformulate problem $\mathcal{P}_1$ as
 \begin{subequations}
  \label{theta simple}
  \begin{align}(\mathcal{P}_3)\,\,\,\,\,\,&\max_{\bm{\Theta},t}\,\,\,\,\,\,\,t
   \label{ts:1}\\ 
 \mbox{s.t.}\quad
 &\,\,  \textrm{(\ref{a1}), (\ref{a3}), (\ref{a4}), (\ref{ps:1})}. \nonumber
 \end{align}
 \end{subequations}
As the channel between $U_1$ and $U_2$ is known, the RIS phase in the cooperative transmission time slot can be designed as $\theta^{(2)}_n=\arg(h_{12})-\arg([\mathbf{h}_{1,r}]_n[\mathbf{h}_{r,2}]^{\ast}_n)$ \cite{2020Intelligent}. Hence, we only need to optimize the RIS phase matrix $\mathbf{\Theta}_{(1)}$. We denote $\nu_n=e^{j\theta_n^{(1)}},\forall n \in \mathcal{N}$, $\bm{\nu}=[\nu_1,\ldots,\nu_N]^T$, and $R_0=2^{\frac{a_1+a_2}{\beta}}-1$. With the assistance of $\bm{\nu}$, we obtain 
$\mathbf{h}_{k}^H\mathbf{\Theta}_{(1)}\mathbf{G}\mathbf{p}_i=\mathbf{d}_{k,i}^{H}\bm{\nu}$, where $\mathbf{d}_{k,i}=(\textrm{diag}(\mathbf{h}_{k}^{H})\mathbf{G}\mathbf{p}_{i})^{\ast}\in \mathbb{C}^N$.
To ease notations, we denote $g_{k,i}=\mathbf{g}_{k}^{H}\mathbf{p}_i$. 
We further introduce slack variables
$\bm{\eta}=[\eta_1,\eta_2]$, $\bm{\delta}=[\delta_1,\delta_2]$, $\eta_{c,2}$, and $\delta_{c,2}$. $\bm{\eta}$ and $\bm{\delta}$ denote the SINR  and rate vectors of private streams, respectively. $\eta_{c,2}$ and $\delta_{c,2}$ denote the SINR and rate of the common stream of  $U_2$ in the first time slot. Then, problem $\mathcal{P}_3$ is reformulated as
\begin{subequations}
  \label{theta hard}
  \begin{align}(\mathcal{P}_{3.1})&\,\,\,\,\,\,\max_{\substack{\bm{\nu},\bm{\eta},\bm{\delta},\\
  \eta_{c,2},\delta_{c,2},t}}\,\,\,\,\,\,\,t   \label{th:1}\\ 
 \mbox{s.t.}\quad
 &\,\beta\delta_k+a_k \geq t,\forall k\in \{1,2\},\label{th:2}\\
 &\,\beta\delta_{c,2}+c_2^{(2)}\geq a_1+a_2, \label{th:3}\\
 &\,1+\eta_k-2^{\delta_k}\geq 0,\forall k\in\{1,2\},\label{th:4}\\
  &\,1+\eta_{c,2}-2^{\delta_{c,2}}\geq 0,\label{th:5}\\
  &\,|\nu_{n}|=1, \forall n\in \mathcal{N},\label{th:6}\\
 &\, \frac{\left|\mathbf{d}_{k,k}^H\bm{\nu}+g_{k,k}\right|^2}{\left|\mathbf{d}_{k,i}^H\bm{\nu}+g_{k,i}\right|^2+\sigma_k^2}\geq \eta_k,\forall k\in\{1,2\},i\neq k, \label{th:7}\\
 &\,\frac{\left|\mathbf{d}_{2,0}^H\bm{\nu}+g_{2,0}\right|^2}{\sum_{i=1}^{2}\left|\mathbf{d}_{2,i}^H\bm{\nu}+g_{2,i}\right|^2+\sigma_2^2}\geq \eta_{c,2},\label{th:8}\\
  &\,\frac{\left|\mathbf{d}_{1,0}^H\bm{\nu}+g_{1,0}\right|^2}{\sum_{i=1}^{2}\left|\mathbf{d}_{1,i}^H\bm{\nu}+g_{1,i}\right|^2+\sigma_1^2}\geq R_0. \label{th:9}
  \end{align}
 \end{subequations}
To handle the non-convexity of constraint (\ref{th:6}), we adopt the penalty method to transform  $\mathcal{P}_{3.1}$ into
\begin{subequations}
\label{penalty method}
 \begin{align}
   \max_{\substack{\bm{\nu},\bm{\eta},\bm{\delta},\\
  \eta_{c,2},\delta_{c,2},t}}\,\,\,\,&\,\,\,t+C\sum_{n=1}^{N}(|\nu_n|^2-1)   \label{pm:1}\\ 
 \mbox{s.t.}\quad
 &\,\,|\nu_{n}|\leq 1,\forall n\in \mathcal{N},\label{pm:2}\\
 &\,\,\textrm{(\ref{th:2})--(\ref{th:5}), (\ref{th:7})--(\ref{th:9})},
 \end{align}   
\end{subequations}
where $C$ is a large positive constant. 
The objective function can be  approximated around the point $(\bm{\nu}^{[l]})$ at iteration $l$ by the first-order Taylor approximation of $C\sum_{n=1}^{N}(|\nu_n|^2-1)$, which is given by
\begin{equation}
t+C\sum_{n=1}^{N}\Re \{2(\nu^{[l]}_{n})^{\ast}\nu_n-|\nu_{n}^{[l]}|^2\}.
\end{equation}
To handle the nonconvexity of constraints  (\ref{th:7}) and (\ref{th:8}), we introduce  variables $\bm{\kappa}=[\kappa_1,\kappa_2]$ and $\kappa_{c,2}$, and
constraints (\ref{th:7}) and (\ref{th:8}) are
equivalent to
\begin{equation}
\label{decompose fraction:1}
\begin{aligned}
|\mathbf{d}_{k,i}^{H}\bm{\nu}+g_{k,i}|^2+\sigma_k^2   &\leq\kappa_k ,\forall k\in\{1,2\}, i\neq k,\\\ 
\sum_{i=1}^{2}|\mathbf{d}_{2,i}^{H}\bm{\nu}+g_{2,i}|^2+\sigma_2^2   &\leq\kappa_{c,2}, 
\end{aligned}
\end{equation}
and
\begin{equation}
\label{decompose fraction:2}
\begin{aligned}
|\mathbf{d}_{k,k}^{H}\bm{\nu}+g_{k,k}|^2   &\geq\eta_k\kappa_k ,\forall k\in\{1,2\}\\
& =\frac{1}{4}((\eta_k+\kappa_k)^2-(\eta_k-\kappa_k)^2),\\
|\mathbf{d}_{2,0}^{H}\bm{\nu}+g_{2,0}|^2 &\geq\eta_{c,2}\kappa_{c,2}\\
& =\frac{1}{4}((\eta_{c,2}+\kappa_{c,2})^2-(\eta_{c,2}-\kappa_{c,2})^2).\\
\end{aligned}
\end{equation}
(\ref{decompose fraction:1}) and (\ref{decompose fraction:2}) are still non-convex, we therefore adopt  the first-order Taylor approximation at point $(\bm{\nu}^{[l]},\bm{\eta}_k^{[l]},\bm{\kappa}^{[l]},\eta_{c,2}^{[l]},\kappa_{c,2}^{[l]})$. (\ref{decompose fraction:1}) and (\ref{decompose fraction:2}) become
\begin{equation}
\label{decompose fraction:3}
\begin{aligned}
&2\Re((\mathbf{d}_{k,k}^{H}\bm{\nu}^{[l]}+g_{k,k})^{H}\mathbf{d}_{k,k}^{H}\bm{\nu})-|\mathbf{d}_{k,k}^{H}\bm{\nu}^{[l]}|^2+|g_{k,k}|^2\\
 \geq&\frac{1}{4}((\eta_k+\kappa_k)^2-2(\eta_k^{[l]}-\kappa_k^{[l]})(\eta_k-\kappa_k)\\
& +(\eta_k^{[l]}-\kappa_k^{[l]})^2),\forall k\in\{1,2\},\\
&2\Re((\mathbf{d}_{2,0}^{H}\bm{\nu}^{[l]}+g_{2,0})^{H}\mathbf{d}_{2,0}^{H}\bm{\nu})-|\mathbf{d}_{2,0}^{H}\bm{\nu}^{[l]}|^2+|g_{2,0}|^2 \\
\geq&\frac{1}{4}((\eta_{c,2}+\kappa_{c,2})^2-2(\eta_{c,2}^{[l]}-\kappa_{c,2}^{[l]})(\eta_{c,2}-\kappa_{c,2})\\
&+(\eta_{c,2}-\kappa_{c,2})^2).\\
\end{aligned}
\end{equation}
Similarly, constraint (\ref{th:9}) can be approximated by
\begin{equation}
\label{decompose fraction:4}
    \begin{aligned}
    &2\Re((\mathbf{d}_{1,0}^{H}\bm{\nu}^{[l]}+g_{1,0})^{H}\mathbf{d}_{1,0}^{H}\bm{\nu})-|\mathbf{d}_{1,0}^{H}\bm{\nu}^{[l]}|^2+|g_{1,0}|^2 \\
    \geq&R_0(\sum_{i=1}^{2}|\mathbf{d}_{1,i}\bm{\nu}+g_{1,i}|^2+\sigma_1^2).
    \end{aligned}
\end{equation}
Based on the above approximation methods, problem $\mathcal{P}_3$ is approximated by the following convex problem at iteration $l$.
\begin{subequations}
\label{theta last}
\begin{align}
 \max_{\substack{\bm{\nu},\bm{\eta},\bm{\delta},\\
  \eta_{c,2},\delta_{c,2},\\
  \bm{\kappa},\kappa_{c,2},t}}\,\,\,\,&\,\,\,t\label{tl:1}\\ 
\mbox{s.t.}\quad
 &\,\,\textrm{(\ref{th:2})--(\ref{th:5}), (\ref{pm:2}), (\ref{decompose fraction:1}), (\ref{decompose fraction:3}), (\ref{decompose fraction:4})}.\nonumber  
\end{align}
\end{subequations}
The detailed SCA method to solve $\mathcal{P}_3$ is illustrated in Algorithm 2.

\begin{algorithm}	
 	\textbf{Initialize}: $n\leftarrow0,t^{[l]}\leftarrow0$, $\mathbf{\Theta}^{[l]}_{(1)}$\;
 	\Repeat{$|t^{[l]}-t^{[l-1]}|<\epsilon$}{
 		$l\leftarrow l+1$\;
 Solve  problem (\ref{theta last}) using  $\mathbf{\Theta}^{[l-1]}_{(1)}$ and denote the optimal value of the objective function by $t^{\star}$ and the optimal solutions by $\mathbf{\Theta}_{(1)}^{\star}$\;
 Update $t^{[l]}\leftarrow t^{\star}$, $\mathbf{\Theta}^{[l]}_{(1)}\leftarrow \mathbf{\Theta}^{\star}_{(1)}$\;			 	}	
\caption{RIS phase optimization algorithm }
\end{algorithm}
\vspace{-2mm}
\begin{algorithm}	
 	\textbf{Initialize}:
 	$n\leftarrow0,t^{[l]}\leftarrow0$, $\mathbf{{P}}^{[l]},\beta^{[l]}, \mathbf{a}^{[l]},\bm{\Theta}^{[l]}_{(1)}$\;
 	\Repeat{$|t^{[l]}-t^{[l-1]}|<\epsilon$} 
 	{$ l\leftarrow l+1$\;
 	  Given $\bm{\Theta}^{[l-1]}_{(1)}$, solve beamforming design  problem using Algorithm 1 and the solution is denoted by $(\mathbf{P}^{[l]},\mathbf{a}^{[l]},\beta^{[l]})$\; 
 	  Given $(\mathbf{P}^{[l]},\mathbf{a}^{[l]},\beta^{[l]})$, solve the phase optimization problem using Algorithm 2  and the solution is denoted by $\bm{\Theta}^{[l]}_{(1)}$ and denote the objective value by $t^{[l]}$\; }
\caption{AO algorithm for problem $\mathcal{P}_1$}
\label{AO algorithm}		
 \end{algorithm} 
 \vspace{-2mm}
\subsection{Alternative Optimization}
 The proposed AO algorithm to jointly optimize the  RIS phases, transmit beamforming, common rate and time slot allocation is shown in Algorithm \ref{AO algorithm}. Starting from a feasible beamforming matrix $\mathbf{P}^{[0]}$, time slot allocation $\beta^{[0]}$, rate allocation $\mathbf{a}^{[0]}$, a RIS phase matrix $\mathbf{\Theta}^{[0]}_{(1)}$, in $l$-th  iteration, we first update  $(\mathbf{P}^{[l-1]},\beta^{[l-1]},\mathbf{a}^{[l-1]})$ with a fixed $\mathbf{\Theta}^{[l-1]}_{(1)}$ by Algorithm 1. For a given $(\mathbf{P}^{[l]},\beta^{[l]},\mathbf{a}^{[l]})$, the RIS phase matrix $\mathbf{\Theta}^{[l]}_{(1)}$ is then updated based on the SCA method with $\mathbf{\Theta}^{[l-1]}_{(1)}$ as the initialization point. The max-min rate $t$ is  then
calculated based on the updated $(\mathbf{P}^{[l]},\beta^{[l]},\mathbf{a}^{[l]})$ and $\mathbf{\Theta}^{[l]}$.  The process is repeated until convergence.
\par 
\textit{Convergence Analysis}: For the precoder optimization problem $\mathcal{P}_2$ , the proposed SCA algorithm ensures the monotonic increase of the objective function. This is due to the fact that the solution of problem (\ref{p-last}) at iteration $l-1$ is a feasible point of problem (\ref{p-last}) at iteration $l$. Due to the transmit power constraint (\ref{a5}), the solution sequence $\{t^{[l]}\}^{l=\infty}_{l=1}$ is bounded above, which implies  that the convergence of Algorithm 1 is guaranteed. Similarly, for the RIS phase problem $\mathcal{P}_3$, the solution of problem (\ref{theta last}) at iteration $l-1$ is also a feasible point for the problem (\ref{theta last}) at iteration $l$. Due to the modulus constraint (\ref{a4}), the solution sequence $\{t^{[l]}\}^{l=\infty}_{l=1}$ is bounded above, which guarantees the convergence of Algorithm 2.  The monotonic increase of the objective functions for $\mathcal{P}_2$ and  $\mathcal{P}_3$ implies that the solution of  $\mathcal{P}_1$  at iteration $l-1$ is also a feasible point for  $\mathcal{P}_1$ at iteration $l$. Hence, the  convergence of AO algorithm is guaranteed.
\section{Numerical Results}
In this section, we evaluate the performance of the proposed system model. 
The following schemes are compared:
\begin{itemize}
\item \textbf{RIS CRS}---This is the  scheme proposed in Section \ref{sec: system model}.
\item \textbf{RIS RSMA}---This is a special case of ``RIS CRS'' when $\beta$ is fixed to 1.
\item \textbf{RIS SDMA}---This is a  special case of ``RIS CRS'' when  the power allocated to the common stream
is zero.
\item \textbf{no RIS CRS}---This  is the CRS scheme without using RIS, as studied in \cite{jian2019crs,Mao2020}.
\item \textbf{no RIS RSMA}---This  is the traditional RSMA scheme without using RIS, as studied in \cite{mao2017rate,RSintro16bruno}.
\item \textbf{no RIS SDMA}---This  is the  multi-user linearly precoded SDMA without using RIS, as studied in \cite{mao2017rate}.
\end{itemize}

The max-min rate  problems of  RIS CRS, RIS RSMA and RIS SDMA are solved by Algorithm 3 while the corresponding problems of no RIS CRS, no RIS RSMA and no RIS SDMA are solved by SCA directly.
\begin{figure}[tb]
\centerline{\includegraphics[width=0.35\textwidth]{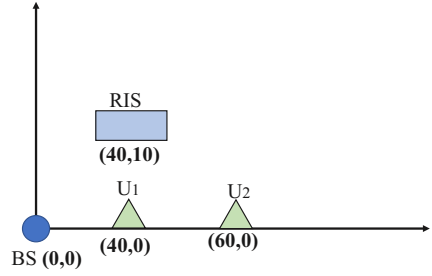}}
\caption{The simulated RIS-aided cooperative rate splitting  scenario.}
 \vspace{-2mm}
\label{fig2}
\end{figure}

\par
The simulation  setting  is shown in Fig. \ref{fig2}. The BS and RIS are located at $(0, 0)$ and $(40,10)$, respectively. $U_1$ and $U_2$ are located at $(40, 0)$ and $(60,0)$. For simplicity, the small-scale fading of all channels are modeled as Rayleigh fading.  
The path loss of the channels are modeled by $P(d)=L_0d^{-\alpha}$, where $L_0=-30$ dB is the  path loss at the reference distance $d_0=1$ m, $d$ denotes the link distance and $\alpha$ refers to the path loss exponent. In particular, the path loss exponents of BS to $U_1$, BS to $U_2$, BS to RIS, RIS to all users and $U_1$ to $U_2$ are set to be $2,3,3,3.5$ and $1.5$. Without loss of generality, we assume the transmit power at the BS and the relaying user $U_1$ are equal.  The noise tolerance is $\sigma^2=-66$ dBm and the convergence tolerance is $\epsilon=10^{-3}$.
  
\begin{figure}[tb]
\centerline{\includegraphics[width=0.35\textwidth]{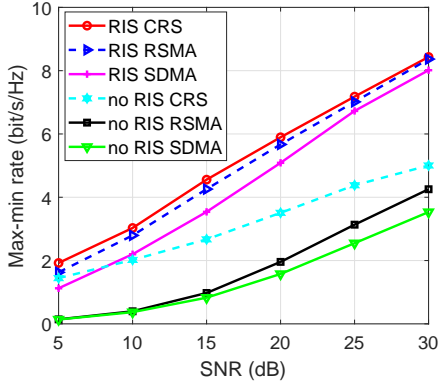}}
\caption{Max-min rate versus the transmitter SNR, when $N_t=2,N=4$.}
\label{fig3}
\end{figure}

Fig. \ref{fig3} shows the max-min rate of different strategies versus the transmit power when $N_t=2,N=4$. It shows
that the proposed  RIS-aided CRS scheme outperforms all other schemes. The relative max-min rate gain of  RIS CRS over  RIS RSMA and no RIS CRS are at least 6.6$\%$ and 19.5$\%$ when SNR is 15 dB. By using the RIS-aided CRS model, the max-min rate of the system  increases significantly.
\par
\begin{figure}[tb]
\centerline{\includegraphics[width=0.35\textwidth]{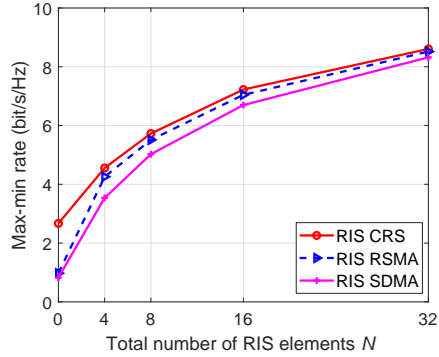}}
\caption{Max-min rate versus the number of RIS Elements, when $N_t=2$ and SNR is 15 dB.}
\label{fig4}
 \vspace{-2mm}
\end{figure}

 
Fig. \ref{fig4} shows the max-min rate of different strategies versus the number of passive reflecting elements at the RIS when the number of transmit antenna is $N_t=2$ and SNR is 15 dB. The three schemes all achieve significant max-min rate improvement as the number of RIS element increases. 
The proposed RIS-aided CRS scheme achieves a higher max-min rate than RIS-aided RSMA and RIS-aided SDMA. But the rate gain  of RIS-aided CRS over RIS-aided RSMA  decreases as the number of RIS elements increases.

\par
\begin{figure}[tb]
\centerline{\includegraphics[width=0.35\textwidth]{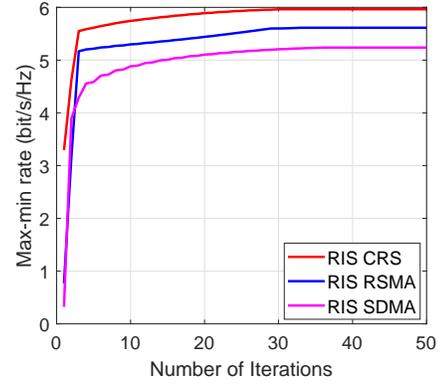}}
\caption{Convergence of the algorithms in one channel realization.}
\label{fig5}
 \vspace{-2mm}
\end{figure}

 
Fig. \ref{fig5} illustrates the convergence of the proposed AO algorithm when $N_t=2, N=8$ and SNR is 15 dB. In general, the algorithm can converge with 50 iterations.
\section{Conclusion}
In this work, we propose an  RIS-aided CRS downlink transmission network. The transmit beamforming vector, the RIS phase matrix, the common rate and time slot allocation are jointly optimized to maximize the minimum rate among users. To solve this problem, we propose an AO algorithm that alternatively optimizes the phase matrix and the remaining variables. Numerical results show the proposed scheme which combines CRS and RIS enhances the spectral efficiency and user fairness. 

\bibliographystyle{IEEEtran}  
\bibliography{reference}

\end{document}